\documentclass[12pt,preprint]{aastex}
\newcommand\nts{\negthinspace\negthinspace\negthinspace\negthinspace\negthinspace\negthinspace}

\begin{document}

\title{X-ray Emission from the Weak-lined T Tauri Binary System KH~15D}

\author{William Herbst and Edward C. Moran}
\affil{Astronomy Department, Wesleyan University, Middletown, CT
06459}
\email{bill@astro.wesleyan.edu,ecm@astro.wesleyan.edu}

\begin{abstract}

The unique eclipsing, weak-lined T Tauri star KH~15D has been detected as an
X-ray source in a 95.7 ks exposure from the {\it Chandra X-ray Observatory\/}
archives. A maximum X-ray luminosity of $1.5 \times 10^{29}$ erg~s$^{-1}$ is
derived in the 0.5--8 keV band, corresponding to $L_{\rm X}/L_{\rm bol} =
7.5 \times 10^{-5}$. Comparison with samples of stars of similar effective
temperature in NGC 2264 and in the Orion Nebula Cluster shows that this is
about an order of magnitude low for a typical star of its mass and age.  We
argue that the relatively low luminosity cannot be attributed to absorption
along the line of sight but implies a real deficiency in X-ray production.
Possible causes for this are considered in the context of a recently proposed
eccentric binary model for KH 15D. In particular, we note that the visible component rotates rather slowly for a weak-lined T Tauri star and has possibly been pseudosynchronized by tidal interaction with the primary near periastron.     

\end{abstract}
 

\section{Introduction}

KH 15D is a unique eclipsing pre-main sequence (PMS) system near the Cone nebula
in NGC 2264 \citep{kehm,kh98}. The visible star is of K6 or K7 spectral class
\citep{ham01,agol} and has an H$\alpha$ equivalent width of $\sim$ 2 \AA,
typical of a weak-lined T Tauri star (WTTS). Its mass and age are $\sim$0.6
$M_\odot$ and 2 My, respectively \citep{ham01}. At high spectral resolution
the star reveals broad wings on its hydrogen emission lines and during eclipse
one clearly sees forbidden emission lines \citep{ham03}. These features signify
that accretion and outflow are still active in the system, but not at the level
of a typical classical T Tauri star (CTTS). While the star has no  measured
infrared excess, an apparent disk and jet in H$_2$ have been detected by
\citet{dem} and \citet{tok}, respectively. 

The extremely long duration of the eclipse, currently about one-half of the
period, clearly shows that the eclipsing body is not a companion star. Rather,
it appears to be part of a circumstellar or circumbinary disk \citep{h02}.
During eclipse the system becomes both
bluer and more highly polarized \citep{h02, agol}, suggesting that we are seeing it primarily or entirely in scattered light.  There are two time scales
associated with the eclipse, a 48.37 day cycle for the main eclipse and a
secular increase in the eclipse duration of about 1 day per year
\citep{h02,w03,ham04a}. Two recent models based on the historic light curve
\citep{w03,jw04} have proposed that the 48 day eclipse cycle is the orbital
period of a binary system while the secular variation is caused by  precession
of the circumbinary disk \citep{w04,c04}. If this is correct it means that,
for the first time, we can probe the structure of a disk on length scales as
small or smaller than a stellar diameter and monitor events in a possibly
planet-forming disk on human time scales! Clearly it is important to understand
as much as possible about this unique PMS stellar system and to
exploit its fortuitous geometry while the opportunity lasts.

One characteristic of T Tauri stars, especially WTTS, is that they are
prodigious sources of X-ray emission, although for still largely unknown
reasons \citep{fm, f03}. We hoped to use the periodic eclipse of the K6-7
star behind an optically thick and presumably X-ray opaque circumstellar disk,
to allow us to map the structure of the coronal plasma in this WTTS. As a
prelude to this intended study we searched for archival X-ray data on NGC
2264 and found a long exposure in the archives of the {\it Chandra X-ray
Observatory\/} that includes KH 15D. It was obtained during a time interval
when the star was out of eclipse so we expected a relatively strong signal,
characteristic of a WTTS. Instead, we found that the total X-ray count out of
eclipse is so small that it may not be possible to learn much by monitoring
during an eclipse cycle. This in turn has prompted us to consider the extent
to which KH 15D is unusual in yet another way, namely as a remarkably faint
X-ray source for a WTTS. In this paper we present the case that it is, indeed,
an unusually weak source of X-ray emission and discuss possible implications
of this for the system and for the broader question of X-ray production in 
solar-like, PMS stars.

\section {X-ray Data}

\subsection{The X-ray Luminosity of KH 15D}

The southern portion of NGC 2264, containing KH 15D, was observed with the ACIS-I array on board the {\it Chandra X-ray
Observatory\/} for 95.74 ks on 2002 October 28--29 (UT).  The instrument was
operated in its nominal mode, with a frame time of 3.2~s and a focal-plane
temperature of --120 $^{\circ}$C.  We reprocessed the data using the CIAO
software, version 3.2.1, to correct for charge-transfer inefficiency of the
front-illuminated CCDs and to apply a time-dependent gain correction to the
data.  In addition, we improved the original afterglow correction and removed
the 0.5-pixel randomization applied during the standard processing of the data.
The screened data used for analysis consist of events with grades 0, 2, 3, 4,
and 6 and energies between 0.5 and 8.0 keV.

The J2000 coordinates of the instrument aimpoint for the observation are 
$\alpha$ = $06^{\rm h} 40^{\rm m} 58.\!\!^{\rm s}$10,
$\delta$ = $+09^{\circ} 34' 00.\!\!{''}40$; thus, KH 15D, at
$\alpha$ = $06^{\rm h} 41^{\rm m} 10.\!\!^{\rm s}$27,
$\delta$ = $+09^{\circ} 28' 33.\!\!{''}40$, is included within the
$17' \times 17'$ ACIS-I field of view at an off-axis angle of 6\farcm35.
A weak source at the location of KH 15D is faintly visible in the image.
We used bright nearby sources to determine the appropriate size of the
source aperture (12 pixels, or $\sim 6''$, in radius), which was centered
on the optical position of KH 15D.  The background level was estimated in
a concentric, source-free annulus with inner and outer radii of 20 and 60
pixels, respectively.  A total of 22.5 net counts were detected in the
full 0.5--8.0 keV band, corresponding to a signal-to-noise ratio of 3.5.
The spectrum of KH 15D is soft, with $\sim 80$\% of the net counts falling
below 2 keV.  The implied band ratio (i.e., the ratio of counts detected in
the hard 2--8 keV and soft 0.5--2 keV ranges) is 0.264, although, because
KH 15D is not formally detected in the hard band, this should be considered
an upper limit.

We used the XSPEC software, along with the response matrix and effective area
files generated by CIAO, to estimate the X-ray flux of KH 15D.  A spectral
model consisting of a single-temperature, optically thin (MEKAL) plasma was
adopted \citep{f02}.  Solar abundances were assumed for the plasma, and
based on the maximum possible reddening of KH 15D of $E(B-V) = 0.1$
\citep{ham01}, an absorption column density of $2 \times 10^{20}$
atoms~cm$^{-2}$ was included in the model.  We adjusted the temperature of
the plasma until the above band ratio was obtained, which suggests $kT = 2.7$
keV.  Using the model normalization required to match the observed count rate
of the source, we obtain a 0.5--8 keV flux of $2.2 \times 10^{-15}$ erg
cm$^{-2}$ s$^{-1}$.  For a distance of 760~pc \citep{sbl,ham01}, this
corresponds to an unabsorbed X-ray luminosity of $1.5 \times 10^{29}$ erg
s$^{-1}$.  Given that the maximum values of the band ratio and column density
were assumed, this should be taken as an upper limit.

\subsection{Comparison with Other PMS Stars}

Our derived total X-ray luminosity for KH 15D may be compared with other PMS stars of similar effective temperature in NGC 2264 and in the Orion Nebula Cluster (ONC). The most direct comparison is with stars on the same archival {\it Chandra} image as KH 15D. We used the photometric and spectroscopic surveys of \citet{r02} and \citet{l04a} to search for K stars within that field that had comparable magnitude (within 0.7 mag in $I$) and color (within 0.04 mag in $V-I$). Since the reddening in NGC 2264 is relatively small (E(B-V) $\sim0.1$ or less) and uniform, this procedure should result in a reasonable comparison set. Five stars meeting the photometric conditions were found and all five were firmly detected in the
{\it Chandra\/} image.  Their source counts and X-ray spectra were extracted
in the same manner as described above for KH 15D.  

Because of the strength
of the detections of the comparison stars, we were able to estimate their X-ray fluxes 
directly via spectral modeling.  We began by fitting single-temperature
plasma models to each of the spectra.  However, good fits were obtained in
just two of the cases.  For the remaining three objects, we employed
two-temperature plasma models, which are frequently required to fit the
X-ray spectra of PMS stars \citep{f02}.  The fitted column
densities for four of the five objects are consistent with zero.  The fit of
other object, star 3748, suggests a column density of $\sim 4 \times 10^{21}$
cm$^{-2}$.  This absorption was not corrected for when calculating the star's
X-ray flux, so the X-ray luminosity we derive for it is a lower limit.

Table \ref{table1} summarizes the optical and X-ray properties of the five
comparison stars and KH 15D.  Listed for each object are the source number
from \citet{r02}, the $I$-band magnitude, the $V-I$ color, the spectral type,
the net counts detected in the 0.5--8 keV band, the hard-to-soft band counts
ratio $H/S$, the plasma temperature(s) in keV, the 0.5--8 keV flux (in erg
cm$^{-2}$ s$^{-1}$), and the X-ray luminosity (in erg s$^{-1}$) in the same
band.  As the table indicates, KH 15D is significantly underluminous relative to this set of
comparison stars.  Moreover, it is evident that the weakness of its X-ray
emission is {\it not\/} a result of a greater amount of soft X-ray absorption:
the upper limit to its band ratio and the inferred plasma temperature are
consistent with those of the comparison stars. We also calculated X-ray fluxes for the ONC stars using exactly the same spectral model as for KH15D, with resultant fluxes that
differed by only 3 to 25\% (with an average deviation of 17\%) from what is shown in Table \ref{table1}.

\begin{figure}
\plotone{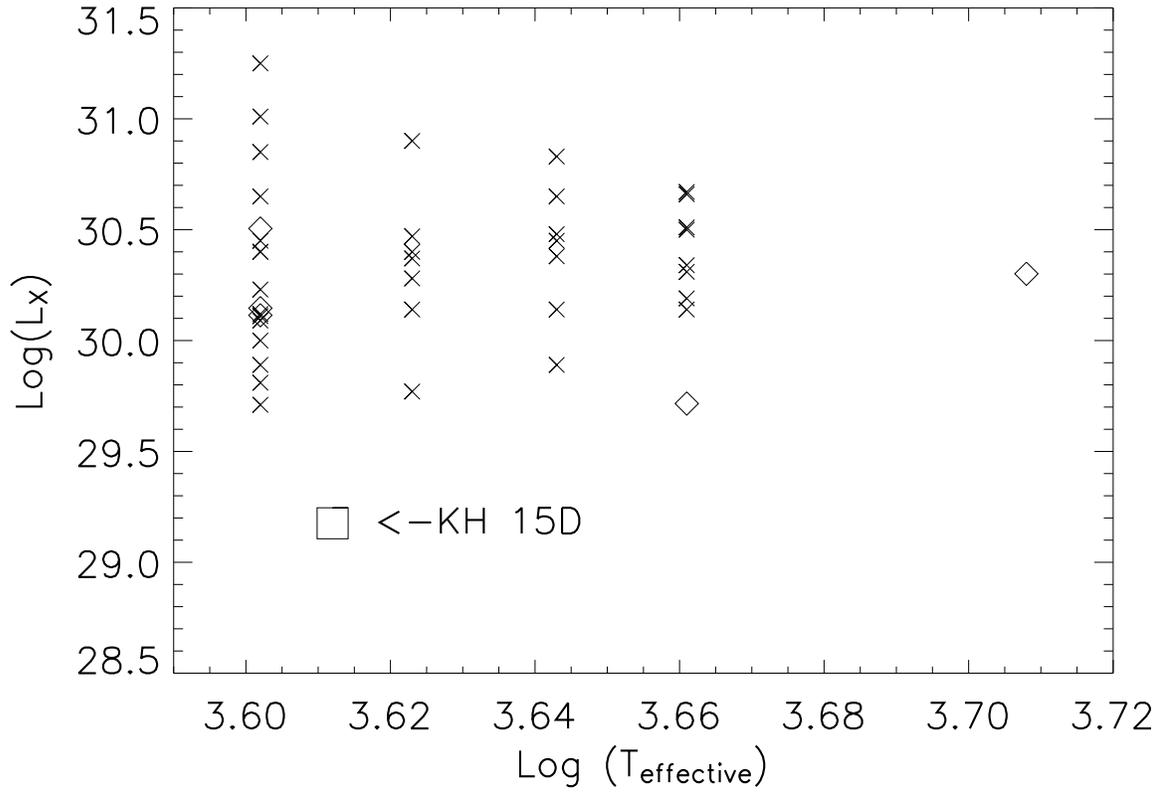}
\figcaption{X-ray luminosity of mid-K spectral class members of the northern part of NGC 2264 from \citet{rrs04} (x's) and from Table 1 (diamonds) compared with KH 15D (square).  Obviously, KH 15D is about an order of magnitude fainter in the X-ray than is typical of cluster members of similar effective temperature.  \label{Fig. 1}}
\end{figure}

To expand the comparison set, we have also employed observations of K3-K7 stars in the northern part of NGC 2264 whose X-ray luminosities were calculated by \citet{rrs04} based on an ACIS-I {\it Chandra} image. They detected 37 likely cluster members in this spectral range, as well as 5 non-members \citep{rs05}, and their exposure time of 48.1 ks is long enough to have reached sources close to the luminosity of KH 15D, if not below it. The X-ray luminosity of their cluster members is derived in a manner similar to what we have used and is based on the same assumed distance. In Fig. \ref{Fig. 1} we compare the X-ray luminosity of KH 15D (large square) to the the stars in the northern part of the cluster (crosses) and to those in the southern part from Table \ref{table1} (diamonds). It is clear from this comparison that KH 15D has a lower than typical X-ray luminosity for mid-K stars in both parts of NGC 2264. It lies about an order of magnitude below the median value of both samples and is more than a factor of three fainter than the next faintest detected object. 

\begin{figure}
\plotone{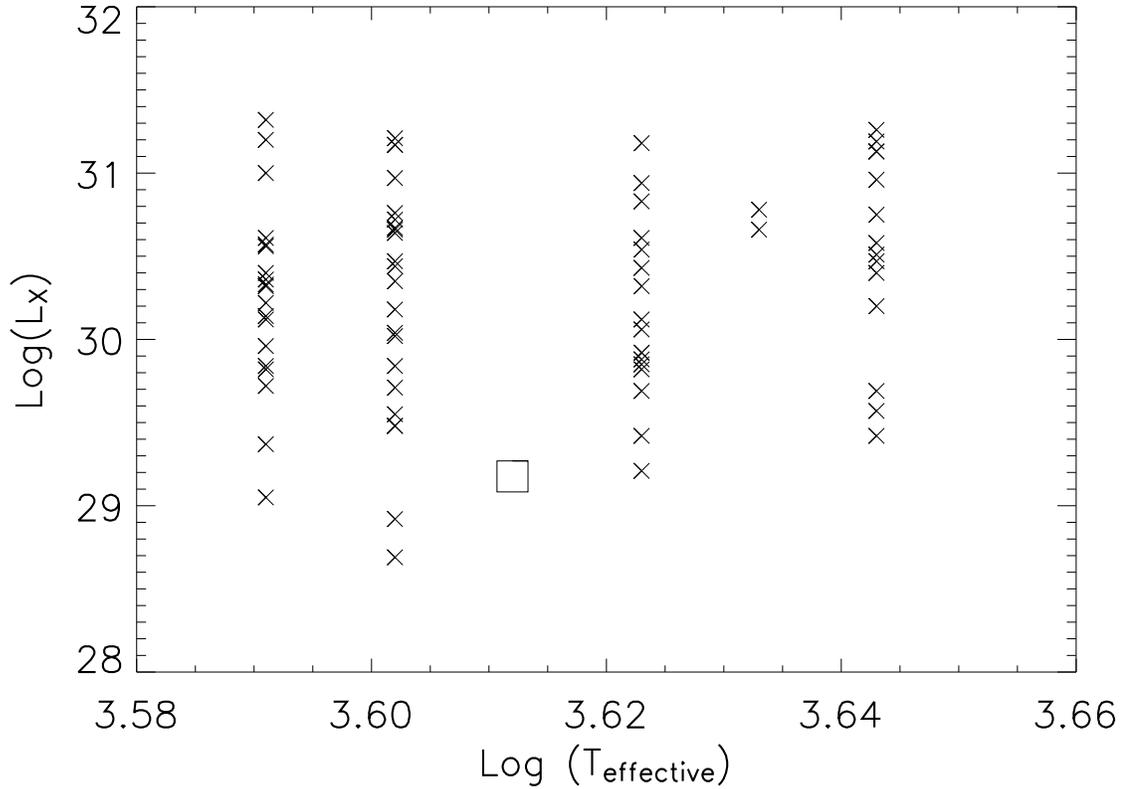}
\figcaption{The X-ray luminosity of KH 15D (square) is compared to the K5--K8 stars in the ONC (x's) from Feigelson et al. (2004).  Extinction corrections have been made for the Orion data and are negligible in the case of KH 15D, as discussed in the text. It is clear that KH 15D is deficient in its X-ray production by about an order of magnitude compared to stars of similar effective temperature in the ONC. Note that two of the three stars in the ONC with smaller X-ray luminosity than KH 15D also have visual extinctions exceeding 5 magnitudes!  \label{Fig. 2}}
\end{figure}

As a further test of the degree to which KH 15D is anomalously weak in X-rays, we compare it to stars of similar spectral class in the ONC. This cluster is slightly younger and about half the distance of NGC 2264. The optical data come from the extensive photometric and spectroscopic survey by \citet{h97}.  Total X-ray luminosities in the {\it Chandra\/} band of 0.5--8 keV have been derived by
\citet{f04} for more than a thousand sources in the ONC, based on an
extraordinarily long exposure of 850 ks. Fig. \ref{Fig. 2} shows a comparison of our result for KH 15D with the 74 ONC members
having spectral types between K5 and K8, inclusive. The inferred plasma temperatures of
these stars are comparable to the value of 2.7 keV that we infer for KH 15D, which is
typical of PMS stars in general \citep{f02}. 
Note that this X-ray survey
detected every optically known  ONC member in this spectral range, so the comparison
sample is as complete as possible.  It
is clear from this figure that KH 15D is a very weak X-ray source compared to
the ONC stars of similar effective temperature. 
 
We further note that KH 15D is likely to be even more anomalously weak as an X-ray emitter than is evident from  Fig. \ref{Fig. 2}. There are two reasons for this. First, the extinction in the ONC is much higher in general than in NGC 2264 and is also highly variable. For example, two of the three ONC stars in Fig. \ref{Fig. 2} with cited X-ray luminosity less than KH 15D have visual extinction estimates exceeding 5 magnitudes! This suggests to us that some of the apparent scatter in X-ray luminosity in the cluster, especially the scatter to low values, is due to extinction, not lack of X-ray production. Second, we note that the ONC is slightly younger and probably has, therefore, a higher percentage of CTTS compared to WTTS. This is hard to verify because of the strong nebulosity in which the ONC is embedded. If true, however, it means that there may be more low luminosity X-ray sources in the ONC owing to this difference.

To summarize this section, we find that KH 15D is the weakest X-ray emitter known for stars of its spectral class in NGC 2264 and lies about 1 order of magnitude below the median value for the cluster. It is also a weaker X-ray source than all but 3 of the 74 known mid-K star members of the ONC, and two of them have visual extinctions exceeding 5 magnitudes!  Again, it lies about an order of magnitude below the median of the cluster. It is possible, of course, that the {\it Chandra\/} exposure we analyzed was obtained, by chance, at a time when KH 15D was at or near the bottom of its range of X-ray variability. It is believed that much of the scatter seen in PMS X-ray luminosity is caused by actual time variations associated with flaring \citep{fm, f03}. This would be an extraordinarily unlucky circumstance, of course, since it is such an extreme outlier, and we consider instead, in the remainder of the paper, whether some aspect of the properties of this star which makes it unique in other ways could also account for its unusually low X-ray luminosity. 

\section{Discussion}

Can the low X-ray luminosity of KH 15D be attributed to some sort of extinction effect, perhaps within the circumbinary disk which surrounds it? We find this difficult to support for two reasons. First, there is essentially no reddening or obscuration evident in the light of the K6--7 star during maximum brightness, when the X-ray data were obtained. The star has a color excess of $E(B-V) = 0.1$ mag if it is a K6 star and less if it is K7. This is consistent with what is found for other members of NGC 2264, in which the reddening is known to be small \citep{r02, l04a}. If there is any local extinction associated with circumstellar matter, it must be very small. Out of eclipse the star also shows very small photometric variations (less that 0.1 mag in $I$) and no detectable color variations \citep{ham04b}. Also, there is no evidence in the X-ray data for a deficiency of soft X-rays which would be most susceptible to absorption. The hard-to-soft ratio is typical of what is found for lightly reddened T Tauri stars such as those in the ONC. We conclude that KH 15D is almost certainly an intrinsically weak X-ray source because of an anomalously low production rate, not because of absorption. 

Since the cause of X-ray emission in PMS stars is not fully established, it is not immediately evident how to interpret the low luminosity of KH 15D. Here we discuss two possible explanations, neither of which is without difficulty. It seems likely that, in some way, the binary nature of the star is an important element, so we begin there. An attractive unifying paradigm for the unique and, in some cases, anomalous properties of this WTTS is provided by the eccentric binary model of \citet{w04}. It is shown by these authors that constraints on the system from the historic and current light curves can be understood if KH 15D is a roughly equal luminosity binary system with a highly eccentric ($e \sim 0.5-0.8$) orbit and period of 48.4 days. The orbit is, at present, slightly inclined to the plane of a circumbinary disk so that one (and only one) of the stars periodically rises above it. Precession of the disk plane is plausibly responsible for the secular variation in the eclipse duration. This compelling model implies a separation of the components at periastron of only about 0.08 AU, close enough to consider possible tidal effects or other influences that such a close approach could have on the system and, in particular, X-ray production.

\subsection{Interacting Magnetospheres at Periastron?}

The radius of the only currently visible star is about 1.3 $R_\odot$ based on its luminosity and effective temperature \citep{ham04b}. The currently invisible companion was last seen in 1995 and measured to be brighter by several tenths of a magnitude than the K7 star. The historical light curve of the system also demonstrates that the unseen companion is slightly more luminous than the visible star. Assuming the stars are coeval, which seems inescapable, simple theoretical considerations demand that the unseen star be slightly more massive and larger than the K7 star. Hence its radius is probably a little larger than 1.3 $R_\odot$ but not much larger. The separation of the two components at periastron is about 15 stellar radii. Since magnetospheres of WTTS are typically believed to extend 5--10 stellar radii from the surfaces \citep{os95, p05} disruption of the magnetosphere by interactions with matter (or magnetic fields) inside of this point could play a role in lessening X-ray luminosity either by cooling or by lack of confinement of hot gas. 

Unfortunately there is little evidence to support (or refute) this hypothesis in the observed X-ray luminosity  of other PMS binaries. Perhaps the most similar system known is DQ Tau, which is a CTTS with nearly equal-mass mid-K components in an orbit of eccentricity e=0.56, which brings the stars within about 8 stellar radii of each other at periastron \citep{msb97}. The star is not detected in the ROSAT All-Sky Survey \citep{kns01}, which means it is weaker than many PMS stars in Taurus. However, since it is a CTTS it is possible that its low X-ray luminosity is due to absorption in circumstellar matter.  Another reasonably eccentric (e=0.24) PMS binary is UZ Tau E \citep{mmd05}. Unfortunately, it is only a few arc-seconds from UZ Tau W (also a binary) and so the X-ray luminosity of this binary is not measured. Its separation at periastron is also a little larger (about 25 solar radii), it has a much smaller mass ratio (q=0.2) and it is also a CTTS, so there are potentially important differences with KH 15D. 

One WTTS spectroscopic binary with a K7 primary, circular orbit and separation of 12.6 stellar radii, V826 Tau, is observed to be roughly normal in its X-ray luminosity, with a quiescent luminosity of around $2 \times 10^{30}$ erg s$^{-1}$ \citep{rln90,cfk96}. This shows that, proximity of stars, by itself, may not be sufficient to disrupt X-ray emission. However, it may be the variation of the magnetic influence caused by an eccentric orbit that is the key to disrupting a dynamo, so V826 Tau may also not be the best analogue. Since there is no observational evidence that proximity of magnetospheres is a sufficient cause to reduce X-ray emission, we consider another aspect of close periastron passages, namely tidal interactions and possible rotational synchronization. 

\subsection{Tidally Influenced Rotation?}

Rotation is a factor in the X-ray luminosity of stars as young as 30 My but it has not been proven to be important in T Tauri stars; evidence to date suggests it is not. Several authors find no correlation between rotation and X-ray emission for PMS stars in the ONC \citep{gc94, f03}. Perhaps the large and variable extinction effects as well as the difficulty of discriminating between WTTS and CTTS in that cluster cause problems with the interpretation. In this regard it will be interesting to see what studies in slightly older and less highly obscured regions such as the Orion Flanking Fields and NGC 2264 will reveal about the role of rotation in X-ray production \citep{rrs04}. Since X-ray emission in TTS is not yet fully understood and since rotation could be a factor in at least some stars, we inquire whether the rotation of the visible star in the KH 15D system is unusual in any way.

There are two methods for determining the rotation rate of a WTTS and both have been employed in the case of KH 15D. Most directly, one can search for periodic fluctuations in the stellar brightness associated with the rotation of a spotted surface. This has been done by \citet{ham04b} and they have detected two clearly significant peaks in the periodogram of out-of-eclipse data at two separate epochs. In both cases, the period was 9.6 days, strongly suggesting that this is, in fact, the rotation period of the visible component of the binary. Confirmation of that comes from a new measurement of v sin i, based on high resolution spectra taken out of eclipse at the Keck and McDonald observatories by the same group. \citet{ham04b} find a value of $v$ sin $i$ = 6.9 $\pm$ 0.3 km s$^{-1}$ (replacing an earlier estimate of v sin i $<$ 5 km s$^{-1}$ by \citet{ham03} that did not take proper account of macroscopic turbulence in the comparison star). Combining the new v sin i measurement with the known radius of the star, $R = 1.3\, R_\odot$, yields an expected rotation period ($P$) of $P$ sin $i$ = $9.4 \pm 0.3$ days \citep{ham01,ham03}. Since sin $i \sim 1$ for this eclipsing system, we conclude that KH 15D has a rotation period of 9.6 $\pm$ 0.1 days. 

This rotation period of KH 15D is somewhat long for a WTTS of its mass in NGC 2264, where the (bi)modal values are near 1 and 4 days \citep{l04a}. It is not, however, the slowest rotator in the cluster. Of the 184 stars with R-I $<$ 1.84 (roughly corresponding to mass $>$ 0.25 M$_\odot$) in the study by \citet{l04b} 22 (12\%) have periods of 9.6 days or longer. If no correlation between rotation period and X-ray emission exists among NGC 2264 stars in general then the significance of KH 15D's slower than usual rotation for the problem discussed here, is obscure. We note, however, that because it is a member of a relatively close binary, the rotation of the visible component may have been affected by tidal interaction with its primary and could be tidally synchronized (or, rather, pseudosynchronization) as discussed by \citet{ham04b}. 

\citet{p81} has shown that the pseudosynchronzation angular rotation frequency is a nearly constant fraction (f) of the orbital angular frequency at periastron for orbits in the eccentricity range, e = 0.3 to 0.8. One may write, therefore, that $$P_{ps} = {{P_{orb} \over f}{{{(1-e^2)}^{3 \over 2}} \over {{(1+e)}^2}}}.$$ Identifying P$_{ps}$ as the measured rotation period, P$_{orb}$ as the orbital period and taking f=0.81, as appropriate to the plausible eccentricity range of KH 15D \citep{jmh04}. we find the equation is satisfied for e = 0.65 $\pm$ 0.01. This is slightly outside the range of solutions (e = 0.68 to 0.8) favored by \citet{jmh04} based on astrophysical grounds (primarily the system's total  mass and mass ratio), but well within the plausible range based on the radial velocity curve. It is also consistent with the best fit eccentricity, that comes from modeling the historical and modern light curves \citep{w05}. An estimate of the time scale for pseudosynchronization based on the work of \citet{z77} suggests that this could have been achieved within a couple of My, as would be required by the inferred age of the KH 15D system.    
 
To summarize, we have found that KH 15D is an unusual stellar system in a new way --- it is a very weak source of X-ray emission for its mass and age. It seems likely to us that the eccentric binary nature and close periastron approach are probably involved in this. One possible mechanism is disruption of the magnetosphere of the visible star (and probably both stars) during repeated periastron passages due to magnetic reconnection events. Another, perhaps more likely, possibility is disruption of the usual magnetic dynamo through tidal interactions which could also be implicated in the slower than normal rotation of the visible component.  Of course, these are not mutually exclusive mechanisms. Further observations are needed to establish the degree to which KH 15D is in fact anomalous in its X-ray properties for a WTTS and whether either proposed mechanism, or perhaps both, can indeed account for the dearth of X-rays emission.

\acknowledgements

We are deeply indebted to the referee, John Stauffer, for his detailed and enormously helpful report on the original manuscript including providing some of the comparison data for Figure 1. We are likewise indebted to Luisa Rebull for her part in making that comparison sample available to us. We thank Mike Simon, Bob Mathieu, Eric Jensen, Soeren Meibom and Josh Winn for helpful suggestions related to binary PMS stars. We thank Eric Feigelson for helpful discussions and for his leadership of the COUP survey which produced the important ONC comparison sample of Fig. 2. This material is partly based on work supported by the National Aeronautics and Space Administration under Grant NAG5-12502 issued through the Origins of Solar Systems Program.

\clearpage

\begin{deluxetable}{ccccccccc}
\tabletypesize{\scriptsize}
\tablecaption{ Optical and X-ray Properties of K1--K7 PMS Stars in NGC 2264\label{table1}}
\tablewidth{0pt}
\tablehead
{
\colhead{star}   & 
\colhead{$I$}   &
\colhead{$V-I$} &
\colhead{SpT} &
\colhead{counts} &
\colhead{$H/S$} &
\colhead{$kT$} &
\colhead{$F_{\rm X}$} &
\colhead{$L_{\rm X}$} 

}

\startdata

3748  & 14.4 & 1.64 & K1  &  192.1 & 1.261 & $> 10$ &  $3.0 \times 10^{-14}$ &
$2.0 \times 10^{30}$ \\

3778 &  13.8  & 1.61 & K7  &  503.6 & 0.106 & 0.6/2.7 & $4.6 \times 10^{-14}$ &
$3.2 \times 10^{30}$ \\

5143  & 13.8  &1.56 & K4 &   114.3 & 0.141 & 1.8  &    $7.5 \times 10^{-15}$ &
$5.2 \times 10^{29}$ \\

5274  & 14.3 & 1.63 & K4  &  256.3 & 0.089 & 0.7/3.0  & $2.0 \times 10^{-14}$ &
$1.4 \times 10^{30}$ \\

5653  & 14.2 & 1.64 & K7  &  211.0&  0.060 & 0.6/3.7 & $1.9 \times 10^{-14}$ &
$1.3 \times 10^{30}$ \\

KH 15D & 14.5 & 1.60  & K6-7 & ~22.5 & \nts$<0.264$ & 2.7   &   $2.2 \times 10^{-15}$ &
$1.5 \times 10^{29}$ \\
\enddata

\end{deluxetable}

\clearpage

\begin{thebibliography}{}

\bibitem[Attridge \& Herbst(1992)]{ah92} Attridge \& Herbst, W. 1992, 
     \apj 398, L61

\bibitem[Agol et al.(2004)]{agol} Agol, E., Barth, A. J., Wolf, S. \& Charbonneau, D.  2004 \apj\ 600, 781 

\bibitem[Carkner et al.(1996)]{cfk96} Carkner, L., Feigelson, E. D., Koyama, K., Montmerle, T. \& Reid, I. N. 1996, \apj\ 464, 286

\bibitem[Chiang \& Murray-Clay(2004)]{c04} Chiang, E. \& Murray-Clay, R. A. 2004 \apj\ 607, 913

\bibitem[Deming, Charbonneau \& Harrington(2004)]{dem} Deming, D., Charbonneau, D. \& Harrington, J. 2004, \apj\ 601, L87 

\bibitem[Feigelson et al.(2002)]{f02} Feigelson, E. D. et al. 2002 \apj\ 574, 258

\bibitem[Feigelson et al.(2003)]{f03} Feigelson, E. D., Gaffney, J. A., Garmire, G., Hillenbrand, L. A. \& Townsley, L. 2003, \apj\ 584, 911 

\bibitem[Feigelson et al.(2005)]{f04} Feigelson, E. D. et al. 2005, in preparation 

\bibitem[Feigelson \& Montmerle(1999)]{fm} Feigelson, E. D. \& Montmerle, T. 1999 \araa\ 37, 363 

\bibitem[Flaccomio et al.(2003)]{fl03a} Flaccomio, E., Damiani, F., Micela, G. et al. 2003, \apj\ 582,398 

\bibitem[Flaccomio, Micela \& Sciortino(2003)]{fl03b} Flaccomio, E., Micela, G., \& Sciortino, S.  2003, \aap\ 397, 611 

\bibitem[Gagne \& Caillault(1994)]{gc94} Gagne, M. \& Caillault, J.-P. 1994, \apj\ 437, 361

\bibitem[Hamilton(2004)]{ham04a} Hamilton, C. M. 2004, Ph. D. thesis, Wesleyan University 

\bibitem[Hamilton et al.(2005)]{ham04b} Hamilton, C. M.  et al. 2005, \aj\ , in preparation 

\bibitem[Hamilton et al.(2001)]{ham01} Hamilton, C. M., Herbst, W., 
Shih, C. \& Ferro, A. J. 2001, \apj\ 554, L201 

\bibitem[Hamilton et al.(2003)]{ham03} Hamilton, C. M., Herbst, W., Mundt, R., Bailer-Jones, C. A. L., \& Johns-Krull, C. M. 2003, \apj\ 591, L45

\bibitem[Herbst et al.(2002)]{h02} Herbst, W. et al. 2002, \pasp\ 114, 1167

\bibitem[Hillenbrand(1997)]{h97} Hillenbrand, L. A. 1997, \aj\ 113, 1733

\bibitem[Hut(1981)]{p81} Hut, P. 1981, \aap\ 99, 126 

\bibitem[Johnson \& Winn(2004)]{jw04} Johnson, J. A. \& Winn, J. N. 2004, \aj\ 127, 2344

\bibitem[Johnson et al.(2004)]{jmh04} Johnson, J. A., Marcy, G. W., Hamilton, C. M., Herbst, W.  \& Johns-Krull, C. M.  2004, \aj\ 128, 1265

\bibitem[Kearns et al.(1997)]{kehm} Kearns, K. E, Eaton, N. L., Herbst, 
W. \& Mazzurco, C. J. 1997, \aj\ 114, 1098

\bibitem[Kearns and Herbst(1998)]{kh98} Kearns, K. E, \& Herbst, W. 
1998, \aj\ 116, 261

\bibitem[Ko\"nig, Neuha\"user, \& Stelzer(2001)]{kns01} Ko\"nig, B., Neuha\"user, R., Stelzer, B. 2001, \aap\ 369, 971

\bibitem[Lamm et al.(2004)]{l04a} Lamm, M. H., Mundt, R., Bailer-Jones, C. A. L., and Herbst, W. 2004, \aap\  417, 557

\bibitem[Lamm et al.(2005)]{l04b} Lamm, M. H., Mundt, R., Bailer-Jones, C. A. L., and Herbst, W. 2004, \aap\  430, 1005

\bibitem[Martin et al.(2005)]{mmd05} Martin, E. L., Magazzu, A., Delfosse, X. \& Mathieu, R. D. 2005, \aap 429, 939

\bibitem[Mathieu et al.(1997)]{msb97} Mathieu, R. D., Stassun, K., Basri, G., Jensen, E. L. N., Johns-Krull, C. M., Valenti, J. A. \& Hartmann, L. W. 1997, \aj\ 113, 1841

\bibitem[Ostriker \& Shu(1995)]{os95} Ostriker, E. \& Shu, F. 1995 \apj\ 447, 813

\bibitem[Park et al.(2000)]{psbk} Park, B., Sung, H., Bessell, M.S., and Kang, Y. 2000, 
\aj\ 120, 894

\bibitem[Preibisch et al.(2005)]{p05} Preibisch, T. et al. 2005, in press

\bibitem[Ramirez et al.(2004)]{rrs04} Ramirez, S. V., Rebull, L., Stauffer, J., Hearty, T/. Hillenbrand, L. A., Jones, B., Makidon, R., Pravdo, S., Strom, S., \& Werner, M. 2004, \aj 127, 2659

\bibitem[Rebull et al.(2002)]{r02} Rebull, L. M., Makidon, R. B., Strom, S. E., Hillenbrand, L. A., Birmingham, A., Patten, B. M., Jones, B. F., Yagi, H. \& Adams, A. T. 2002, \aj\ 123, 1528

\bibitem[Rebull \& Stauffer(2005)]{rs05} Rebull, L. M. \& Stauffer, J. 2005, private communication 

\bibitem[Reipurth et al.(1990)]{rln90} Reipurth, B., Lindgren, H., Nordstrom, B. \& Mayor, M. 1990, \aap\ 235, 197

\bibitem[Sung, Bessel and Lee(1997)]{sbl} Sung, H., Bessell, M. S. \& Lee, 
S-W. 1997, \aj\ 114, 2644

\bibitem[Tokunaga et al.(2004)]{tok} Tokunaga, A. T. et al. 2004, \apj\ 601, L91 

\bibitem[Winn et al.(2003)]{w03} Winn, J. N., Garnavich, P. M., Stanek, K. Z. \& Sasselov, D. D. 2003 \apj\ 593, L121

\bibitem[Winn et al.(2004)]{w04} Winn, J. N., Holman, M. J., Johnson, J. A., Stanek, K. Z., \& Garnavich, P. M. 2004, preprint.

\bibitem[Winn et al.(2005)]{w05} Winn, J. N. et al., in preparation

\bibitem[Zahn(1977)]{z77} Zahn, J. -P.  1977, \aap  57, 383

\end{thebibliography}
\end{document}